# Dissipative Magneto-optic Solitons

A. D. Boardman, L. Velasco and P. Egan

*Joule Physics Laboratory, Institute for Materials Research, University of Salford,
Salford, Manchester,M5 4WT, United Kingdom*
`a.d.boardman@salford.ac.uk`

## 1  Introduction

The study of magneto-optics involves the polarization state of light [1-4], which is a measure of its vector nature. The displacement vector in a magneto-optic medium is $\mathbf{D} = \varepsilon_0 [\boldsymbol{\varepsilon} \cdot \mathbf{E} + i\mathbf{g} \times \mathbf{E}]$, where $\boldsymbol{\varepsilon}$ is the relative permittivity that exists in the absence of an applied magnetic field, $\varepsilon_0$ is the permittivity of the free space, $\mathbf{g} \propto f(\mathbf{H}_{app})$ is called the gyration vector, and the function $f(\mathbf{H}_{app})$ involves the applied magnetic field $\mathbf{H}_{app}$. The vector $\mathbf{g} \times \mathbf{E}$ is normal to $\mathbf{E}$ and the overall dielectric property tensor of magneto-optic material has off-diagonal elements. For magneto-optic phenomena, as opposed to chiral properties, $\mathbf{g}$ does not depend upon which way the light wave is travelling. This is an important qualification because it means that it continues to act in the same direction, even after the wave has been forced to change direction by a reflection. Hence the physical property described by the form of $\mathbf{D}$ is non-reciprocal. A further generalization is that $\mathbf{g}$ may depend upon the spatial coordinates. It is common practice to write the displacement vector as $\mathbf{D} = \varepsilon_0 \boldsymbol{\varepsilon} \cdot \mathbf{E}$, where $\boldsymbol{\varepsilon}$ is now the complete magneto-optic tensor. It is also common practice to write the off-diagonal terms of $\boldsymbol{\varepsilon}$ as $\pm iQn^2$, where $n$ is the refractive index of the un-magnetised dielectric and $Q(\mathbf{r})$ is called the magneto-optic parameter distribution. Some common magneto-optic configurations are shown in Figure 1, in which the direction of the saturation magnetisation $\mathbf{M}$, relative to the propagation direction, is shown.

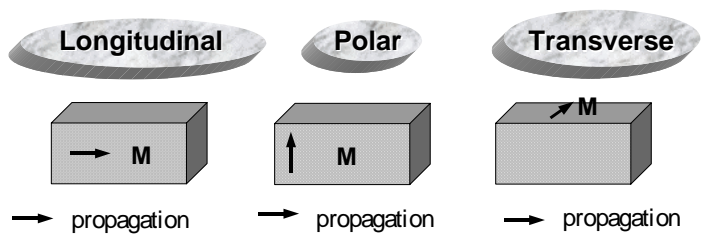

**Figure 1 Common magneto-optic configurations.** $M$ **is the magnetization.**

The longitudinal case is usually called the Faraday configuration. In the bulk this will cause a rotation about the propagation direction of the electric field carried by a plane wave propagating along $M$. This propagation can be resolved into the propagation of two counter-rotating circularly polarised waves, each seeing a different refractive index. A reversal of the propagation direction reverses these indices and non-reciprocal behaviour occurs. The transverse case maybe called either the Voigt or Cotton-Mouton configuration. This is reciprocal in the bulk and non-reciprocal in an asymmetric waveguide.

Magneto-optics was once described as the stepchild [5] of integrated optics [2]. This impression originated from the once simple desire to "insert" magneto-optics into known designs rather than address and control the fascinating complexity of the materials. A major task, for example, is using third-order optically nonlinear materials interfaced to magneto-optic materials so that bright solitons can be controlled. This is made possible by the fact that magneto-optic materials have been brought to a high state of readiness by the magnetism community, because the latter maintains a very strong interest in magneto-optic recording media and optically non-reciprocal devices [6]. These include periodic structures and ultra-thin films, the general aim being to exploit modern controllable magnetic properties. Magneto-optic behaviour, however, is a specific, non-reciprocal example of gyrotropic behaviour. In fact the term gyrotropic embraces optical [1-4] activity and general chiral properties [7] but it is the non-reciprocal behaviour of magneto-optics that is so important for applications. As has been eloquently stated before [8], the vector nature of light leads to robustness and high discrimination levels in applications. The combination of polarisation and nonlinearity is therefore a very powerful mix. Indeed, when coupled to photo-induced Faraday rotation it is possible to discriminate this effect from the background of other nonlinear effects [9]. Non-reciprocal behaviour is characteristic of artificial gyrotropy [2] and can be used in optical isolators and a range of coherence and quantum problems. This is all in sharp contrast to natural gyrotropy, like optical activity. This has been highlighted as being tuned to non-local field-matter interactions and so can be used to probe chiral molecular systems [9]. Finally, it is important to go beyond third-order, Kerr, optical nonlinearity and move towards a saturable model of terms in the polarisation. In this spirit, this chapter seeks to determinate the influence of a magneto-optic presence upon an optically nonlinear material that is modelled by a cubic-quintic form of polarisation. In addition, the coefficients of the envelope equation will be made complex, to take into account both linear and nonlinear damping and cubic gain processes. The emphasis is upon the simulation outcomes, however, rather the applications.

## 2   The basic cubic-quintic complex Ginzburg-Landau equation

The cubic-quintic complex Ginzburg-Landau (CQGLE) equation derives its character not only from an extension of the widely used cubic Kerr nonlinearity to include a quintic contribution, but also from the inclusion of complex coefficients that model loss and gain. The action taken to broaden the scope of the nonlinearity is a step towards acknowledging that the nonlinearity of many materials saturate, as the propagating power increases. The generalisation to complex coefficients permits the modelling of gain and losses, both of a linear and a nonlinear origin. Adding in a magneto-optic material property will modify the basic envelope equation in a rich variety of ways, depending upon whether the full vector character of the propagating electromagnetic waves is simulated.

Optical beams can diffract but this tendency is, broadly speaking, offset by the ability of the material nonlinearity to self-focus the beams. A *scalar* electric field, $E(x,z,t)$, associated with the beam satisfies the standard wave equation

$$\frac{\partial^2 E}{\partial z^2} + \nabla_\perp^2 E + \varepsilon(\omega)\frac{\omega^2}{c^2}E = 0 \tag{1.1}$$

where $\nabla_\perp^2 = \frac{\partial^2}{\partial x^2} + \frac{\partial^2}{\partial y^2}$, and the fast time dependence of the electric field varies as $e^{(-i\omega t)}$ where t is time and $\omega$ frequency. The velocity of the light in the vacuum is $c$ and $\varepsilon(\omega)$ is the *total* relative permittivity. The spatial dependence of $E$ can be written as $\psi(x,z)e^{(ik_0 z)}$, where $k_0 = \frac{\omega}{c}\sqrt{\varepsilon_L}$ and $\varepsilon_L$ is the linear permittivity. The complex amplitude, $\psi(x,z)$, is actually slowly varying, i.e. $\psi(x,z)$ evolves slowly enough for $\left|\frac{\partial^2 \psi}{\partial z^2}\right| \ll \left|k_o \frac{\partial \psi}{\partial z}\right|$ to be true. This is a perfectly reasonable assumption as any numerical simulation will reveal. The appropriate form of the wave equation is, therefore,

$$2ik_0 \frac{\partial \psi}{\partial z} + \nabla_\perp^2 \psi + \left[\varepsilon(\omega)\frac{\omega^2}{c^2} - k_0^2\right]\psi = 0 \tag{1.2}$$

For a non-dispersive, linear, medium the final term will vanish, since $\varepsilon(\omega) = \varepsilon_L$ and $k_0^2 = \varepsilon_L \frac{\omega^2}{c^2}$. The presence of nonlinearity, creates a small, but significant, nonlinear contribution to the difference $k_o^2 - \varepsilon\frac{\omega^2}{c^2}$. The term $\left(\varepsilon\frac{\omega^2}{c^2} - k_0^2\right)$ is $\frac{k_0^2}{n_0^2}(\sqrt{\varepsilon} + \sqrt{\varepsilon_L})(\sqrt{\varepsilon} - \sqrt{\varepsilon_L})$. This is approximately $\frac{k_0^2}{n_0^2}(n_2|\psi|^2)$, after writing the nonlinear index as $n_0^2 + n_2|\psi|^2$, so that $\varepsilon$ is $\varepsilon_L + 2n_2|\psi|^2 n_0$. Hence, to a first approximation, the final term in (1.2) is $\frac{k_o^2}{n_0}(n_2|\psi|^2)\psi$, retaining, for the moment, only cubic nonlinearity. The inclusion of nonlinear effects up to fifth order involves the introduction of another constant, $n_4$, defined through the envelope equation

$$2ik_0 \frac{\partial \psi}{\partial z} + \nabla_\perp^2 \psi + 2\frac{k_o^2}{n_0}n_2|\psi|^2 \psi - 2\frac{k_o^2}{n_0}n_4|\psi|^4 \psi = 0 \tag{1.3}$$

This equation can be usefully scaled by measuring $x$ and $y$ in the units $D_0$, equal to the beam width, and $z$ in Rayleigh, or diffraction, lengths $2k_0 D_0^2$. The transformations to the now dimensionless coordinates and the admission of complex coefficients results in the generation of the familiar cubic quintic equation

$$i\frac{\partial \psi}{\partial z} + i\delta\psi + \left(\frac{1}{2} - i\beta\right)\left(\frac{\partial^2 \psi}{\partial x^2}\right) + \left((1-i\varepsilon)|\psi|^2 - (\nu - i\mu)|\psi|^4\right)\psi = 0 \qquad (1.4)$$

The physical interpretation of (1.4) leads to the conclusion that $\delta$ accounts for any linear absorption, $\beta$ represents diffusion, $\varepsilon$ is the nonlinear cubic gain, $\nu$ is the quintic coefficient that measures the self-defocusing brought on by the negative sign in the last term, and $\mu$ is a nonlinear loss term, of quintic origin.

This is a powerful model that has been shown to generate many interesting solutions [10] through the elegant work of Akhmediev and Ankiewicz [11]. The addition of a magneto-optic influence requires some care [12-15], however, because the choice of applied magnetic field orientation and the choice of the spatial distribution of the magnetisation are both going to be important.

## 3   Magneto-optics with inhomogeneous magnetisation

This section focuses only upon the influence of the magneto-optics and diffraction upon beam envelope behaviour in classical Faraday and Voigt configurations. The outcomes can be legitimately added later to the effects of nonlinearity. The magneto-optic effect is a perturbation that ranks alongside the perturbation that nonlinearity represents and the combined effect, including diffraction, produces a slowly varying evolution in the electric field amplitude. These are the reasons why magneto-optic behaviour can be studied separately in this section without adding in the nonlinearity until later.

Assuming the time dependence $e^{-i\omega t}$, Maxwell's equations for the electric field vector $\mathbf{E}$ lead to

$$\nabla^2 \mathbf{E} + \frac{\omega^2}{c^2}\boldsymbol{\varepsilon} \cdot \mathbf{E} - \nabla(\nabla \cdot \mathbf{E}) = 0 \qquad (1.5)$$

For propagation along the $z$-axis, the permittivity tensor depends upon the direction of an externally applied field $\mathbf{H_0}$. This field produces a magnetisation distribution defined through a function $Q$ that is often used as a constant proportional to the magnitude of the saturation magnetisation. If the applied magnetic field is supplied by a single wire electrode, however, $Q$ is a function of the spatial coordinates. To be specific, if the wire lies along the $z$-axis then the magnetic field created will be a vector that is tangential to circles in the $(x, y)$ plane, centred upon the wire. There are components of the magnetisation parallel to the *x*- and *y*-axes but only the one parallel to the *x*-axis will be important, and $Q \equiv Q(x)$ models such a magnetisation distribution.

The principal configurations modelled are :

**Faraday**:
$$\varepsilon = \begin{pmatrix} n^2 & -iQn^2 & 0 \\ iQn^2 & n^2 & 0 \\ 0 & 0 & n^2 \end{pmatrix} \quad (1.6)$$

**Voigt**:
$$\varepsilon = \begin{pmatrix} n^2 & 0 & 0 \\ 0 & n^2 & -iQn^2 \\ 0 & iQn^2 & n^2 \end{pmatrix} \quad (1.7)$$

where it is assumed, based upon the known properties of most magneto-optic materials, that all the diagonal elements are equal. For a layered structure, this form of $Q(x)$ really is readily provided by a wire electrode arrangement deposited upon the upper planar surface, or buried within the structure. For the Faraday configuration some arrangement of layers, or magnetic domain structure, may be used. Basically, permitting $Q$ to be a function of any the coordinates is a matter for experimental ingenuity on the one hand, but at the same time it is a feature that adds exciting functionality to the behaviour of optical beams.

The key to magneto-optic behaviour is the fact that $\nabla \cdot \boldsymbol{E} \neq 0$, and this is characteristic of a forced gyrotropic medium. The divergence of the total displacement vector is always zero, however, so that

$$\nabla \cdot (\boldsymbol{\varepsilon} \cdot \boldsymbol{E}) = 0 \quad (1.8)$$

*Faraday configuration*

In this case, the full electric field in a guiding structure is $\boldsymbol{E} = (E_x, E_y, E_z) e^{(-i\omega t)}$. For propagation along the *z*-axis, which is also the direction of the applied magnetic field, $E_z \simeq 0$ in a typical planar waveguide and $E_z = 0$, in the bulk. For a simple guiding structure consisting of a cladding-(high-index-core)-substrate arrangement, the validity of neglecting $E_z$ will depend upon the size of the refractive-index change across an interface. The answer is that even though $E_z$ rises quite rapidly with this change, it saturates quickly and remains small, compared to the other electric field components. A strategy of neglecting $E_z$ is usually safe, provided that the core-cladding and the core-substrate materials have refractive indices that match to within 1-5%. The neglect of $E_z$ leads to

$$\nabla \cdot \boldsymbol{E} = i\frac{dQ(x)}{dx}E_y + iQ(x)\frac{\partial E_y}{\partial x} - iQ(x)\frac{\partial E_x}{\partial y} \tag{1.9}$$

Hence, in the Faraday configuration, linear coupled equations emerge which are

$$\nabla^2 E_x + \frac{\omega^2}{c^2}n_m^2 E_x - iQ\frac{\omega^2}{c^2}n_m^2 E_y + F = 0 \tag{1.10}$$

$$\nabla^2 E_y + \frac{\omega^2}{c^2}n_m^2 E_y + iQ\frac{\omega^2}{c^2}n_m^2 E_x + G = 0 \tag{1.11}$$

where

$$F = -iQ\frac{\partial^2 E_y}{\partial x^2} + iQ\frac{\partial^2 E_x}{\partial x \partial y} - iE_y\frac{d^2 Q(x)}{dx^2} - 2i\frac{\partial E_y}{\partial x}\frac{dQ(x)}{dx} + i\frac{dQ(x)}{dx}\frac{\partial E_x}{\partial y} \tag{1.12}$$

$$G = -iQ(x)\frac{\partial^2 E_x}{\partial y^2} + iQ(x)\frac{\partial^2 E_y}{\partial y \partial x} - iQ(x)\frac{\partial^2 E_x}{\partial y^2} \tag{1.13}$$

The presence of a magnetisation distribution would seem, at this stage of the development to be a major complication but a dimensional analysis of the above equations reveals the situation to be otherwise. First, $x$- and $y$-directions can be measured in units of $D_0$, the beam width. In a planar waveguide structure only $x$ will come into play, but both $x$ and $y$ will feature in the bulk applications described later. For guided waves, $E_y$ is carried as a TM wave, and $E_x$ as a TE wave, and they have slightly different wave numbers. An assumption of phase matching is needed for simplification to be exercised but it is not unreasonable in order to remove this 'form' birefringence, and in the bulk this problem does not arise. The $z$-axis can be measured in terms of diffraction, or Rayleigh, length which is $kD_0^2$, where $k$ is the bulk wave number, or the average wave number, if the $E_x$, $E_y$ components are associated with a waveguide. Given this scaling system, the leading terms involving $Q$ have the factor $2kD_0^2 n^2$, but none of the $Q$ terms in $F$ and $G$ have this factor. Typically $Q \sim 10^{-4}$ and $\frac{\omega^2}{c^2}D_0^2 n^2 \sim 10^4$, so that the effective magneto-optic parameter is a transformation of $Q$ to $Q_1 \sim 1$, which is four orders of magnitude more significant than the $Q$ terms in $F$ and $G$. Hence $F$ and $G$ may be neglected.

For the bulk or waveguide structure the electric field components will be written as

$$E_x = \psi_x(x,y,z)\exp\left(i\left(\frac{\omega}{c}\beta\right)\right) \tag{1.14}$$

$$E_y = \psi_y(x,y,z)\exp\left(i\left(\frac{\omega}{c}\beta\right)\right) \tag{1.15}$$

where $\psi_x$, $\psi_y$ are complex and are slowly varying functions of $z$, $\beta = \dfrac{\beta_x + \beta_y}{2}$, and $\beta_x = \beta_y$ in the bulk. The adoption of a rotating coordinate system [13] in which $\psi_+ = \dfrac{1}{\sqrt{2}}(\psi_x + i\psi_y)$, $\psi_- = \dfrac{1}{\sqrt{2}}(\psi_x - i\psi_y)$, where $\psi_+$, $\psi_-$ are appropriately scaled, uncouples (1.10) and (1.11) to give, for example,

$$i\frac{\partial \psi_+}{\partial z} + \frac{1}{2}\left(\frac{\partial^2 \psi_+}{\partial x^2} + \frac{\partial^2 \psi_+}{\partial y^2}\right) \pm Q_1 \psi_+ = 0 \qquad (1.16)$$

This is the equation that gives the development of a circularly polarised wave. It is satisfying that the transformation shows that an arbitrarily polarised input to a Faraday system evolves, in the linear case, as uncoupled counter-rotating, circularly polarised waves. Equation (1.16) will be a description of a guided wave if $\dfrac{\partial^2 \psi_+}{\partial y^2} = 0$, and a beam in a bulk medium if both $x$ and $y$ diffraction terms are retained.

*Voigt configuration*

To use this effect, a guided wave structure is needed. In fact, for a planar structure TM waves are the only modes that can be deployed. They propagate here along the $z$-axis with an electric field [15]

$$\mathbf{E} = \psi\left[\hat{\mathbf{y}}\xi_y(y) + \hat{\mathbf{z}}\xi_z(y)\right]e^{i(\omega t - \beta z)} \qquad (1.17)$$

where $\psi$ is again a slowly varying amplitude, and $\xi_y(y)$, $\xi_z(y)$ are the modal field components. Using the permittivity tensor obtained by (1.7) the wave equation is

$$\nabla(\nabla \cdot \mathbf{E}) - \nabla^2 \mathbf{E} = \frac{\omega^2}{c^2}(\boldsymbol{\varepsilon} \cdot \mathbf{E}) \qquad (1.18)$$

Multiplying (1.18) by $\mathbf{E}^*$ and integrating over $y$ leads to the conclusions that

$$\int \mathbf{E}^* \cdot \nabla(\nabla \cdot \mathbf{E})\,dy = 0 \qquad (1.19)$$

$$\int \mathbf{E}^* \cdot \nabla^2 \mathbf{E}\,dy = \int\left[-\frac{\omega^2}{c^2}n^2|\psi|^2 - 2i\frac{\omega}{c}\beta\psi^*\frac{\partial \psi}{\partial z}\right]\left(|\xi_y|^2 + |\xi_z|^2\right)dy \qquad (1.20)$$

$$\int n^2 \mathbf{E}^* \cdot \mathbf{E} dy = |\psi|^2 \int n^2 \left( |\xi_y|^2 + |\xi_z|^2 \right) dy \qquad (1.21)$$

$$\int \mathbf{E}^* \cdot \mathbf{P}_M dy = -\varepsilon_0 |\psi|^2 \int \left[ \xi_y^* \xi_z - \xi_z^* \xi_y \right] n^2 Q dy \qquad (1.22)$$

where $\mathbf{P}_M$ is the part of the polarisation arising from the magnetisation. After further manipulations equations (1.19), (1.20), (1.21) and (1.22) yield, neglecting diffraction to make sure that the emphasis is upon the magneto-optic effect for this part of the argument,

$$i \frac{\partial \psi}{\partial z} = \frac{\omega}{c} \bar{\varepsilon}_{yz} \psi \qquad (1.23)$$

where $\varepsilon_{yz} = -Q n^2$ and

$$\bar{\varepsilon}_{yz} = \frac{c}{\omega \beta^2} \frac{\int \varepsilon_{yz} \xi_y \left( \frac{\partial \xi_y}{\partial y} \right) dy}{\int \left( |\xi_y|^2 + |\xi_z|^2 \right) dy} \qquad (1.24)$$

This integration over a planar waveguide structure vanishes if a symmetric guide is used. A convenient asymmetric system consists of a magneto-optic layer sandwiched between dissimilar substrate and cladding material.

For an application of (1.23) an asymmetric guiding planar structure will be used, in which the layers will be of infinite extent in the *x*-direction but any optical beams will be restrained by guiding in the *y*-direction. This is why the modal fields are needed to average the magneto-optic effect over the system. Adding diffraction of the beam to (1.23) is achieved in a straightforward manner simply by including the $\frac{\partial^2 \psi}{\partial^2 x}$ term.

Optical nonlinearity can be *added* into (1.16) and (1.23), in cubic-quintic form, without having to be concerned with any cross-phase modulation terms. Adopting the Rayleigh length to measure distances along the *z*-axis, and the beam width as the unit of measurement for distances along the *x*- and *y*-axes creates dimensionless envelope equations. The magneto-optic effect can also be reduced to a dimensionless parameter $Q_1$, and scaling the amplitudes with $\sqrt{L_D/L_{NL}}$, where $L_D$ is the diffraction length and $L_{NL}$ is nonlinear length in the Faraday case, and setting $\psi = \dfrac{c}{\omega D_0}\dfrac{\sqrt{2}}{\sqrt{\beta\chi}}\psi'$ in the Voigt case, where $\chi$ is the cubic nonlinear coefficient averaged over the guiding structure, finally brings the envelope equations into a tidy dimensionless overall form. This action, coupled to the addition of a cubic-quintic nonlinearity, together with complex coefficients, means that the master envelope equations are, replacing $\psi'$ by $\psi$, for convenience:

**FARADAY CONFIGURATION**: bulk medium supporting circularly polarised optical vortex

$$i\frac{\partial \psi}{\partial z} + i\delta\psi + \left(\frac{1}{2} - i\beta\right)\left(\frac{\partial^2 \psi}{\partial x^2} + \frac{\partial^2 \psi}{\partial y^2}\right) + (1 - i\varepsilon)|\psi|^2 \psi - (\nu - i\mu)|\psi|^4 \psi + Q_1(x)\psi = 0 \quad (1.25)$$

**VOIGT CONFIGURATION**: asymmetric planar guiding structure

$$i\frac{\partial \psi}{\partial z} + i\delta\psi + \left(\frac{1}{2} - i\beta\right)\left(\frac{\partial^2 \psi}{\partial x^2}\right) + (1 - i\varepsilon)|\psi|^2 \psi - (\nu - i\mu)|\psi|^4 \psi + Q_1(x)\psi = 0 \quad (1.26)$$

In each system $\delta$ is a measure of the linear damping, $\beta$ represents the possibility of diffusion, $\varepsilon$ is the cubic gain, $\mu$ is a quintic loss and $\nu$ is clearly a self-defocusing contribution to the beam evolution. $Q_1$ is the magneto-optic parameter, which has been made a function of $x$ for the reasons presented earlier.

It must be emphasised that although the length scales are the same in equations (1.25) and (1.26), the amplitudes are scaled slightly differently, but this fact has no impact upon the study of the evolution process. Note also that $Q_1$ is the same order of magnitude in each application but that $Q_1$ in the Voigt case involves a modal averaging factor arising from the guiding in the asymmetric planar structure. This modal factor is the order of unity and will "switch off" $Q_1(x)$ if the waveguide is symmetric. Given the master equations (1.25) and (1.26) it is now possible to explore the behaviour of $\psi$ in each case through extensive simulations.

## 4     Dissipative solitons in Voigt configuration

This section concerns one-dimensional spatial solitons. The beam propagation direction is perpendicular to the direction of an applied magnetic field. Guiding takes place in the *y*-direction and diffraction is permitted in the *x*-direction. The soliton behaviour is controlled by the complex cubic-quintic envelope equation deploying a parameter selection inspired and informed by the work of Akhmediev and his collaborators. For this reason, the computer experiments use a range of parameters that set the conditions $\delta > 0, \beta > 0, \varepsilon > 0, \nu > 0$ and $\mu > 0$ and are given values that permit stable beams to be generated. The input beam for the simulations is an *m*th-order super-Gaussian beam so that for very large values of *m* the cross-section of the beam becomes almost rectangular, and has rather sharply sloping sides. This choice of input beam cross-section creates the impression of the presence of an initial constant background. The intrinsic diffraction-nonlinear chirp balancing, coupled to the imposed gain-loss balance can give substantial additional flexibility through the application of an applied magnetic field. It is difficult, however, to manage the parameters, so, as others have done [10, 16], $\varepsilon$ is selected as a control parameter, whilst keeping the others at fixed values. This is a logical selection, since $\varepsilon$ provides the nonlinear gain of the system.  Also its variance is known to permit discrimination among three main types of evolved stationary soliton beams, so it is interesting to see how the magnetic field impacts upon these beam evolutions, which have been described as plain (in the sense of being ordinary, or what might be expected), composite and moving.  A composite beam is interesting because it retains its flat super-Gaussian shape, but it also acquires a kind of 'hat' or 'cap' that sits on top of this flat region. Clearly this can be technically discussed in terms of source-nonlinear front interaction.

Given this background information, the beam evolutions investigated here deploys the super-Gaussian input beam $\psi(x, z=0) = \exp\left(-\left(\frac{x}{15}\right)^8\right)$, where the order 8 is selected to present a sufficiently wide beam on the input plane. Also $\delta = 0.5, \beta = 0.5$, $\nu = 0.1$ and $\mu = 1.0$, and these parameters are given fixed values, based upon previous experience. For zero applied magnetic field, and $\varepsilon = 2.50$, figure 2(a) shows that a plain beam emerges at $z = 100 L_d$. The selected *z* distance of 100 Rayleigh lengths is considered to be large enough to see the stationary state beam being created. This choice of evolution distance is, of course, only a 'rule of thumb' but it is intuitively acceptable. The stationary beam shape can be understood from the expected tendency of any wide beam to self-focus, or self-trap, until an ordinary, or 'plain' spatial soliton is produced.  For this set of parameters the nonlinear gain is not strong enough to prevent this happening.  A slight change of $\varepsilon$ is enough, however, to produce a composite beam, as is clearly shown in Figure 2(b).

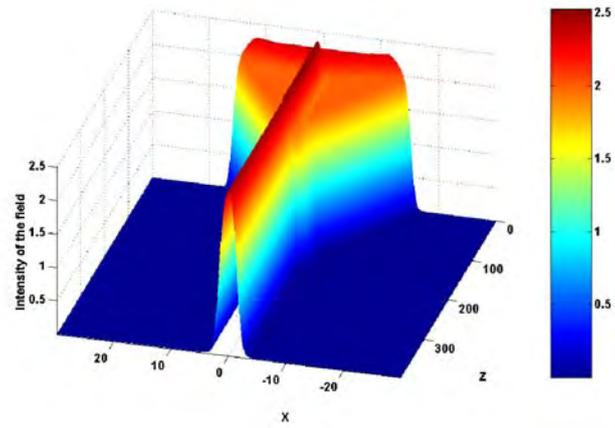

**Figure 2(a) Formation of a plain beam with**
$\beta = 0.5, \; \delta = 0.5, \; \nu = 0.1, \mu = 1, \; \varepsilon = 2.50, \; Q_1(x) = 0$

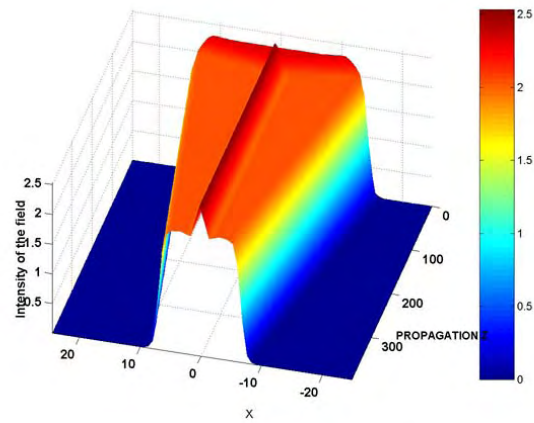

**Figure 2(b) Formation of composite pulse using**
$\beta = 0.5, \; \delta = 0.5, \; \nu = 0.1, \mu = 1, \; \varepsilon = 2.52, \; Q_1(x) = 0$

The optical beams under discussion are contained within a planar layered structure that interfaces nonlinear material to magneto-optic material and is subjected to an applied magnetic field, supplied by placing a current strip onto the upper surface of the waveguide. This arrangement and waveguide type is sketched in Figure 3.

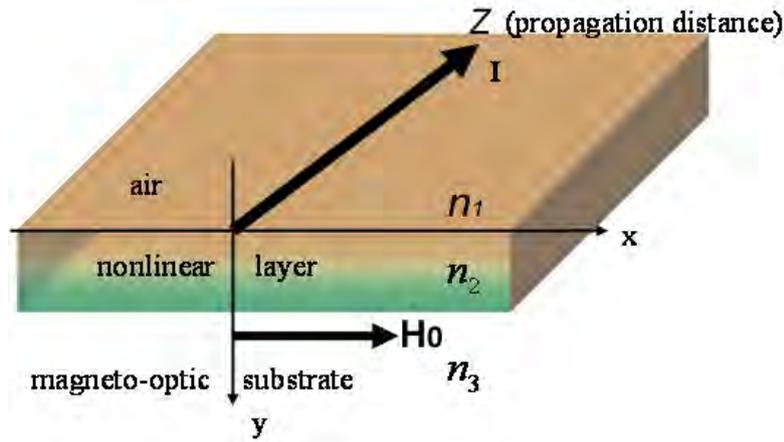

**Figure 3 Voigt configuration waveguide system.**

A single thin wire carrying a current I is shown and the net effect is to provide access to the magnetic field component directed along the *x*-axis. The $n_i$ are the, respective, refractive index values of the guide. Only the *x*-component of the applied magnetic field $\boldsymbol{H_0}$ is shown.

A very thin strip of conducting material, laid down upon the top surface is quite sufficient to supply the current that will have a typical value of 100mA, but can be considerably lower. The strip need only be the order of 20μm wide, so it produces a magnetic field distribution that approximates that of an infinitely thin wire. The magnetic field produced is tangential only directly beneath the wire i.e the magnetic field vector will not in general be parallel to the *x*-axis. The resultant magnetisation will have a component along both the *x*- and y-directions. The component along the *x*-axis gives rise to the transverse, effect but the component long *y* gives rise to a polar effect that can cause TE-TM coupling. No attempt is being made, however, to phase match the birefringence of this guide, so the role of this polar effect is negligible. It is the component of the magnetisation along the *x*-axis that counts. Specifically, if the applied magnetic field at a given coordinate position is at some angle $\phi$ to the *y*-axis, then the total magnetisation will be multiplied by sin($\phi$) to get the *x*-component. This action has the effect of confining the magnetisation to a tight region underneath the wire. Typically, such a region would be approximately -0.25μm ≤ *x* ≤ 0.25μm and these distances are where the magnetisation starts to dip below its saturation value, again, for typical magnet-optic materials. The magnetisation is introduced into the model through a magneto-optic parameter, $Q_1$, and the arguments given above show that it can be a function of *x*. Note that a possible functional shape for $Q_1$ ought to be captured from a straightforward calculation of the magnetization induced by a single current wire. For more complex arrays of wires this task will be more formidable. Even for the simpler case, however, it appears that there is no analytical formula to establish the relationship of the magnetisation to the applied magnetic field. Nevertheless, a simple hyperbolic tangent function is a model that is rather close to experimentally observed dependence of magnetisation upon applied magnetic field. Accordingly, the model $Q_1(x) = A\tanh\left(BH/H_s\right)$ can be used in which A, B are just empirical constants that are selected to make $\tanh \to 1$ in the region of high magnetic field. $H_s$ is the saturation magnetic field and is selected to make sure that $Q_1(x)$ acquires its saturation value. It must be emphasised that $Q_1(x)$ can assume whatever shape is desired, depending upon how the magnetisation is created. For a single current wire lying on top of a planar guide a typical shape of $Q_1(x)$ is sketched in Figure 4. This shows that the magnetisation looks like a 'hill' or 'valley' depending upon the direction of the current and the applied magnetic field direction.

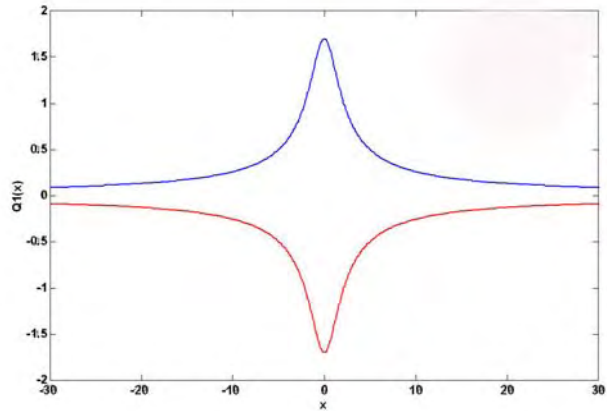

**Figure 4. Typical $Q_1(x)$ distributions for electric currents that flow along +z and –z respectively.**

Returning now to the full complex cubic-quintic Ginzburg-Landau equation, the influence of the $Q_1(x)$ distribution just discussed will be investigated and the parameter set $\beta = 0.5, \delta = 0.5, \nu = 0.1, \mu = 1, \varepsilon = 2.54$ will be adopted. In the absence of any magnetisation, $\varepsilon$ = 2.54 is above the threshold for the total parameter set and causes a super-Gaussian input beam to evolve towards a plane wave. The question to address is whether introducing $Q_1(x)$ maintains this dynamic behaviour, or whether the introduced nonreciprocal property creates a significant change to it. Non-reciprocity is expected because $Q_1$ creates forced gyrotropy, but the extent of this needs to be discovered through simulation. The role of $Q_1$ can be seen in qualitative terms from Equation 1.26. Basically, the nonlinearity is provided by a competition between the cubic self-focusing and the quintic self-defocusing nonlinearity i.e. the function $(|\psi|^2 - \nu|\psi|^4)\psi$, leaving aside the dissipative and gain terms, for the moment. The magneto-optic terms are of the form $\pm Q_1(x)\psi$, so that the plus sign indicates a possible numerical strengthening of the self-focusing term, and the minus sign indicates a strengthening of the self-defocusing term. Admittedly, this is only a qualitative, intuitive, sort of argument but it does indicate the features that should be looked out for in the simulations.

Figure 5 shows how the applied magnetic field influences the soliton development from a super-Gaussian input. Figure 5(a) shows the outcome for $Q_1 = 0$, and sets up the reference behaviour for the parameter set. Figures 5(b) and 5(c) show what happens when the magnetisation is given by

$$Q_1(x) = 1.95 \tanh\left(\frac{I}{2\pi\sqrt{x^2+b^2}} \frac{1}{23.87}\right)$$

where the thickness of the nonlinear guide is $\sim 1\mu m$ and the magnitude of the current I is $\sim 200 mA$. The current is assumed to be positive when flowing along the positive $z$-direction. It can seen in the figure that the qualitative argument is borne out, and that for the backward flowing (along negative z) current the beam rapidly self-focuses, while for the forward (along position z) flowing current the beam rapidly defocuses. Hence a carefully selected $Q_1$ distribution can be used to control the behaviour of the beams in any way that is desired. The behaviour is also clearly non-reciprocal.

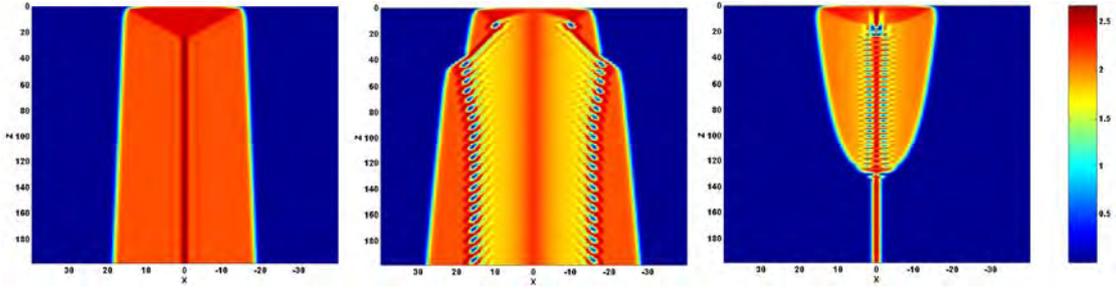

**Figure 5.** $\beta = 0.5$, $\delta = 0.5$, $\nu = 0.1, \mu = 1$, $\varepsilon = 2.53$ **(a)** $Q_1(x) = 0$ **(b)** $Q_1(x) < 0$ **(c)** $Q_1(x) > 0$

The cubic gain and the quintic loss are crucial to the expected stability of any state that evolves from the super-Gaussian input beam. Indeed, it is possible to discover through computer experiments the broad range of $(\mu,\varepsilon)$ values that will lead to a stable evolution in the $Q_1 = 0$ regime. It has been noted in particular that there are some values of $(\mu,\varepsilon)$ that can drive composite *pulses* to be unstable enough [10] to lead to the spontaneous creation of *moving pulses*, where none might be expected. It would be an enormous task to investigate the influence of a whole range of $Q_1 > 0$ or $Q_1 < 0$ distributions upon this conclusion but it is possible to select, as a tool, the type of $Q_1$ used earlier on, and to use sech-type *beams* that are deliberately directed at an angle to the $z$-axis. This is entirely equivalent, mathematically, to pulses that are moving at a finite velocity with respect to their frame of reference. To launch a sech-profiled beam at an angle to the $z$-axis requires the input beam to possess a phase factor so that $\psi(x, z=0) = 2\,\text{sech}(x-x_0)e^{(-i\theta(x-x_0))}$, where $x_0$ is the beam centre, and the proper interpretation of $\theta$ is that $\tan^{-1}(\theta)$ is the angle that the propagation direction of the beam makes to the $z$-axis. Because of the neglect of $\frac{\partial^2 \psi}{\partial z^2}$ in the envelope equation, $\theta$ is small and $x = x_0 + \theta z$. The parameter set selected for the simulations shown in Figure 6(a) is $\beta = 0.5$, $\delta = 0.5$, $\nu = 0.1, \mu = 0.8, \varepsilon = 1.868$, $\theta = 4°$ and the reference calculation is once again the $Q_1 = 0$ case.

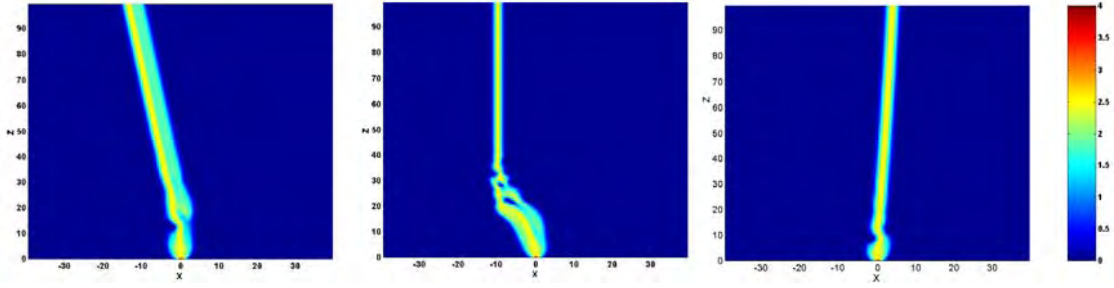

Figure 6(a). Evolution of a beam initially directed at a small angle to the z-axis $\beta = 0.5$, $\delta = 0.1$, $\nu = 0.1, \mu = 0.8$, $\varepsilon = 1.868$, $\theta = 4.75°$ a) $Q_1(x) = 0$; 6(b). Off-axis beam behaviour for $Q_1 > 0$; 6(c): Off-axis beam behaviour for $Q_1 < 0$.

This $Q_1 = 0$ behaviour can be discussed in terms that associate its behaviour [10] with a combination between a front and a beam. The application of an external magnetic field by placing a current wire at $x = -10$ on the upper surface of the waveguide, and causing the current to flow along the $z < 0$ direction, creates the situations illustrated in Figure 6(b) and 6(c), respectively. In one case, the beam remains on the positive side because a potential barrier created by the magneto-optic properties does not allow it to move from there. Also, the beam has been prevented from becoming established in the region that is clearly accessible when $Q_1 = 0$. In the other case, the beam sets itself along the straight line defined by the current wire direction and once again, the beam has been confined. The magnetic field also stops the formation of the "composite beam" that is associated with the non-magnetic case.

To investigate the role of the magnetic field in the intersection dynamics an initial input beam $A(x, z = 0) = 2\operatorname{sech}(x - 10)\exp(-i\theta(x - 10)) + 2\operatorname{sech}(x + 10)$ composed of a beam set an angle to the z-axis and a plain beam, is selected. Each of them is shifted 10 units from the origin and are 20 units apart. Otherwise the parameters are those used above for the single beam.

Figure 7 (a) shows a typical beam intersection for $Q_1 = 0$, in which the beams can coalesce. Note that one of the beams is set at a zero angle to the z-axis and that it remains locked in this position. The introduction of a magneto-optic influence through setting $Q_1 < 0$ or $Q_1 > 0$ creates the effects shown in Figures 7(b) and (c). In the $Q_1 > 0$ case, the effect of the magnetic field is to prevent an interaction between the two beams, and when $Q_1 < 0$ the two beams merge but yet continue to exist in a stationary single beam state. Once again, the magneto-optic term in the envelope equation acts to create a potential that can act as a barrier or an attractive well. This also can be visualised in terms of the quantity $n_2|\psi|^2 - n_4|\psi|^4 \pm Q_1(x)$.

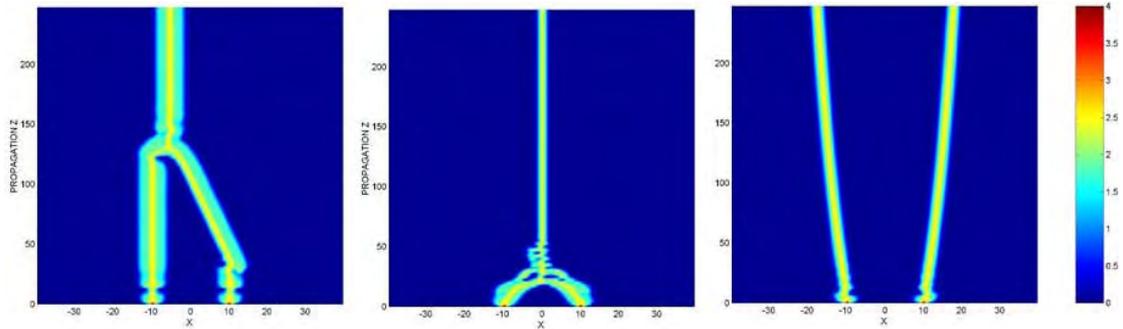

Figure 7 Interacting beams $\beta = .5$, $\delta = 0.1$, $\nu = 0.1, \mu = 0.8$, $\varepsilon = 1.868$ a) Intersection of two beams creating a fused outcome for $Q_1(x) = 0$ b) $Q_1(x) > 0$ c) $Q_1(x) < 0$

## 5 Optical singularities in dissipative media

The study of optical singularities includes the behaviour of vortices [16] and these will now be investigated for a bulk medium, which makes its presence felt through the occurrence of diffraction along *both* the *x*- and *y*-directions. The introduction of magneto-optic behaviour will be through the Faraday effect. This choice is driven by the fact that in the Voigt configuration the magneto-optic effect in bulk media appears only to order $Q_1^2$. After scaling this term to the dimensionless form used in the computations then the dimensionless $Q_1$ used in the waveguide calculations is the order of 1 whilst in the Voigt configuration in the bulk the dimensionless $Q_1$ is the order of $10^{-4}$. In other words it can be neglected. To make the Faraday case simpler to handle only Equation (1.25) will be used which is a description of a circularly polarised beam, arrived at by a transformation of the original Faraday equations to rotating coordinates. The approach to the computations will still be based upon using a broad input beam and numerical experiments that use Equation (1.25), solved with a variable nonlinear gain $\varepsilon$.

A circularly polarised beam, containing two optical vortices of topological charge equal to -1, is modelled at the input plane by

$$A(r, z = 0) = A_0 \left( (x+5) + i(y+5)\operatorname{sgn}(m) \right)^{|m|} \exp\left(-\left(\frac{r_1}{r_0}\right)^2\right)$$

$$+ A_0 \left( (x-5) + i(y-5)\operatorname{sgn}(m) \right)^{|m|} \exp\left(-\left(\frac{r_2}{r_0}\right)^2\right)$$

$$r_1 = \sqrt{(x+5)^2 + (y+5)^2}; \quad r_2 = \sqrt{(x-5)^2 + (y-5)^2}; \quad r_0 = 5; \quad m = -1$$

If there is no magnetic field, the pair of optical vortices attract each other until they combine and form a stable optical vortex, centred at (0,0). Figure 8(a) shows its intensity plot and Figure 8(b) displays the phase and the interferogram. This is an example of the conservation of the angular momentum.

If the magnetic field distribution $Q_1 = 2.34\tanh\left(\dfrac{x}{12}\right)\Big/\cosh\left(\dfrac{x}{6}\right)$ is applied then, for the same propagation distance, the two vortices disappear and the original light beam splits into two smaller diameter ones.(see figure 9). These soliton-like beam are stable, and the magnetic field can control these 'particles'.

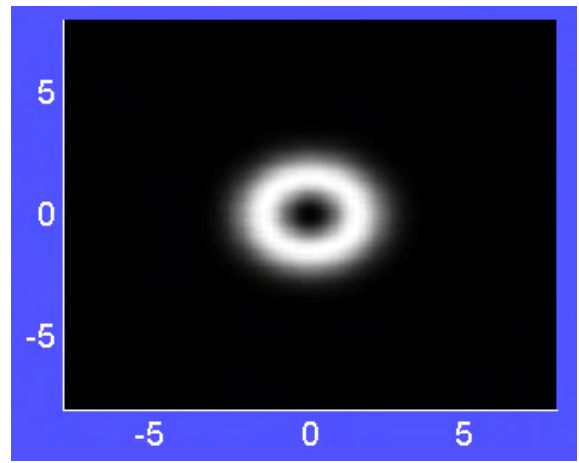

**Figure 8 a) Intensity at $z = 120$ shows the result of the union of the two initial vortices.**

$\varepsilon = 2.5 \;\; \beta = 0.5, \;\; \delta = 0.5, \;\; \nu = 0.1, \;\; \mu = 1, \;\; Q_1(x) = 0$

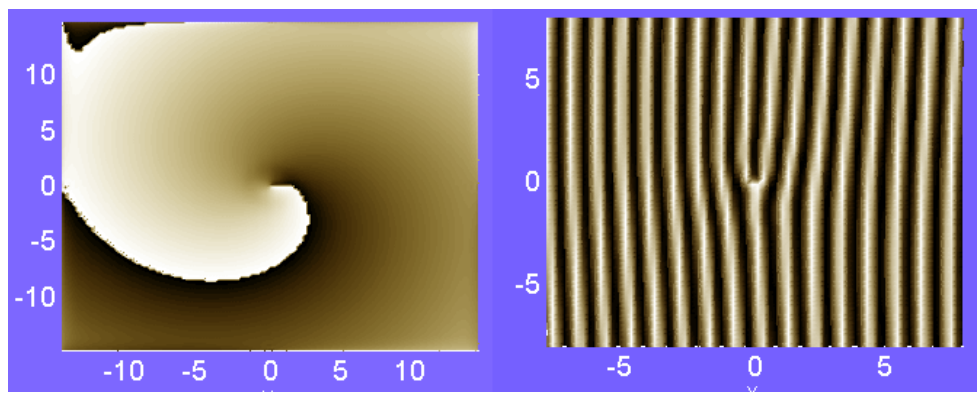

**Figure 8 (b). Phase and interferogram for the merged vortex case.**

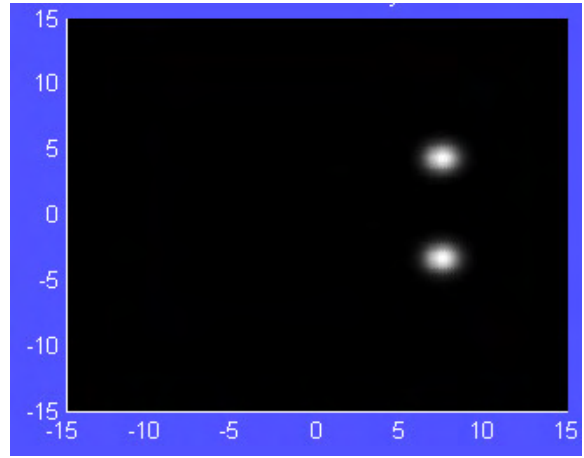

**Figure 9 Propagation to** *z* **= 120 shows that two soliton-like beams develop.**

A simulation using the same parameters but with the vortices slightly further apart at $(6,6)$ and $(-6,-6)$ maintain their integrity, as shown in Figure 10 a). The aim of this simulation is to have the vortices far enough apart to avoid union, as in the case $Q_1 = 0$. Figure 10 b) shows the effect of the magnetic field. This result is similar to that found when the two vortices were close enough to interact. Clearly the magnetic field has two effects. Firstly, it eliminates the singularity of the beam, transforming it into bright soliton-like beams. Secondly, the bright-solitons can be stimulated to move by the action of the magnetic field.

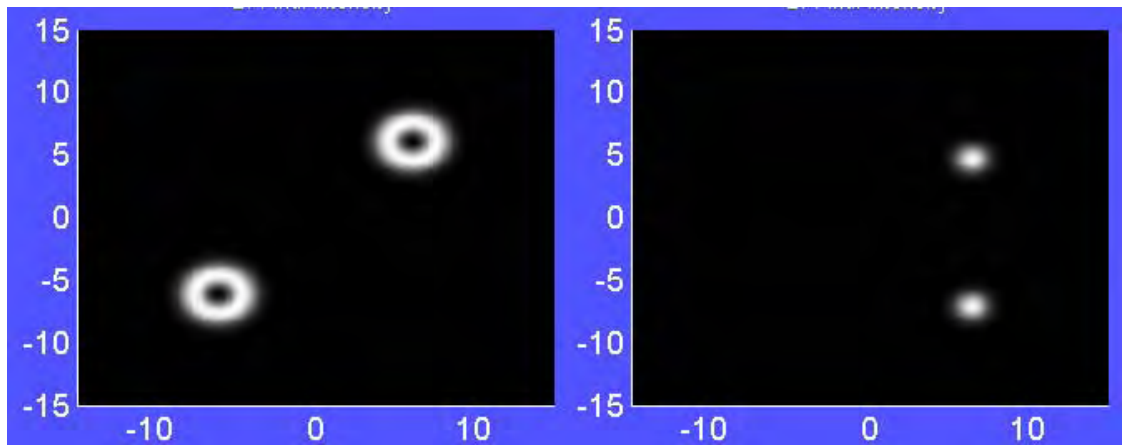

**Figure 10 Intensity at z= 120 , parameters are $\beta = 0.5$, $\delta = 0.5$, $\nu = 0.1$, $\mu = 1$ and $\varepsilon = 2.5$ a) no magnetic field applied, notice the vortex integrity is maintained b)magnetic field is ON**